\documentclass[prb,twocolumn,showpacs]{revtex4}
\usepackage{amsmath}
\usepackage{dcolumn}
\usepackage{graphicx}

\begin{document}

\title[Short title for running header]{$Z_{2}$ spin liquid ground state of the spin-$\frac{1}{2}$ Kagome antiferromagnetic Heisenberg model}
\author{Tao Li}
\affiliation{Department of Physics, Renmin University of China,
Beijing 100872, P.R.China}
\date{\today}

\begin{abstract}
We show that the best variational ground state of the spin-$\frac{1}{2}$ Kagome antiferromagnetic Heisenberg model with nearest-neighboring exchange coupling(NN-KAFH) is a $Z_{2}$ spin liquid state, rather than the widely believed $U(1)$ Dirac spin liquid state. The spinon excitation in the $Z_{2}$ spin liquid state has a small gap of about $1/40$ of the spinon band width. We find that while the $Z_{2}$ and the $U(1)$ spin liquid state have a large overlap on finite clusters and are thus very close in energy, they host totally different spinon excitation spectrum. The strength of the RVB theory becomes particularly clear in such a situation, as it provides not only a variational understanding of the ground state structure, but also a comprehensive picture for the excitation spectrum of the system. Our result indicates that the spin-$\frac{1}{2}$ NN-KAFH should be better understood as a nearly critical system, rather than a prototypical gapped $Z_{2}$ spin liquid system. 
\end{abstract}


\maketitle
The spin-$\frac{1}{2}$ Kagome antiferromagnetic Heisenberg model(KAFH) is an extensively studied system of frustrated quantum magnet in the search of quantum spin liquid\cite{Elser,Chalker,Leung,Young,Lecheminant,Sindzingre,Nakano,Lauchli,Series,Vidal,Singlet,Mila,Auerbach,Poilblanc,Sheng,He1,Changlani}. While it is now generally believed that the ground state of the spin-$\frac{1}{2}$ KAFH with nearest-neighboring exchange coupling(NN-KAFH) is a quantum spin liquid, it is strongly debated on the exact nature of such a novel quantum state of matter. Variational studies based on the resonating valence bond(RVB) theory have accumulated extensive evidences for the $U(1)$ Dirac spin liquid scenario\cite{Hastings,Ran,Iqbal1,Iqbal2,Iqbal3}. This conclusion is also supported by several recent numerical studies with other approaches\cite{Liao,He,Jiang}. On the other hand, a gapped $Z_{2}$ spin liquid state has also been claimed as the true ground state of the system by many other studies\cite{Jiang1,Yan,Depenbrock,Jiang2,Kolley,Gong,Wen}.

According to the RVB theory, a $Z_{2}$ gapped spin liquid state can be realized in the vicinity of the $U(1)$ Dirac spin liquid state when one introduce second-neighboring RVB parameters\cite{Lu,Wen1}. In a previous work of us\cite{Tao1}, we find that a $Z_{2}$ spin liquid state with a small spinon gap is indeed more stable than the widely believed $U(1)$ Dirac spin liquid state. We also find that while the $Z_{2}$ and the $U(1)$ Dirac spin liquid state have a large overlap on finite clusters, and are thus very close in variational energy, they host very different spinon excitation spectrum. This result is challenged by a further variational study\cite{Iqbal4,Iqbal5}, which claims that when the energy difference between the $U(1)$ and the $Z_{2}$ spin liquid states is extrapolated to larger system size, it will eventually become negative. However, as will be shown in this work, such an extrapolation is unreliable since the variational energy of the $U(1)$ and the $Z_{2}$ spin liquid state exhibit rather different scaling behaviors at large system size.

In a recent work\cite{Tao2}, we find that the mapping between the RVB parameters and the spin liquid state is not always injective on the Kagome lattice as a result of a special flat band physics. In particular, we find that the $U(1)$ Dirac spin liquid state with only nearest-neighboring RVB parameter\cite{Hastings,Ran} can be generated from a continuous family of gauge inequivalent mean field ansatzs, whose RVB parameter on the second and the third neighboring bonds are identical. However, the spinon spectrum corresponding to these mean field ansatzs are very different from each other and depend sensitively on the strength of the second neighboring RVB parameter. This finding implies that the RVB parameters encode more information than just the ground state structure. It also implies that the optimization of the RVB parameters around the $U(1)$ Dirac spin liquid state is a rather subtle problem. In particular, one should include the second and the third neighboring RVB parameters simultaneously in the variational description of the spin liquid state of the spin-$\frac{1}{2}$ NN-KAFH. 

Building on these new developments, we have reinvestigated the variational ground state of the spin-$\frac{1}{2}$ KAFH with both the second and the third neighboring RVB parameters. We have performed large scale variational optimization and finite size scaling analysis for the spin-$\frac{1}{2}$ KAFH and find that the best variational ground state of the system is a $Z_{2}$ gapped spin liquid state. The spinon gap in the thermodynamic limit is found to be about $1/40$ of the spinon band width. We find that although the $Z_{2}$ and the $U(1)$ spin liquid state have a rather large overlap on finite clusters, and are thus very close in energy, they host totally different spinon excitation spectrum. Our result demonstrates clearly the strength of the RVB theory, which provides not only an understanding on the ground state structure, but also a comprehensive picture for the excitation spectrum of the system. 

The spin-$\frac{1}{2}$ KAFH studied in this work has the Hamiltonian
\begin{equation}
H=J\sum_{<i,j>}\mathrm{S}_{i}\cdot\mathrm{S}_{j}.\nonumber
\end{equation}
The sum is over nearest-neighboring bonds. To describe the spin liquid ground state of the system in the RVB scheme, we introduce Fermionic slave particle $f_{\alpha}$ and represent the spin operator as $\mathrm{S}=\frac{1}{2}\sum_{\alpha,\beta}f^{\dagger}_{\alpha}\sigma_{\alpha,\beta}f_{\beta}$. Such a representation is exact when the slave Fermion satisfy the constraint $\sum_{\alpha}f^{\dagger}_{\alpha}f_{\alpha}=1$. The RVB state is generated from Gutzwiller projection of the mean field ground state of the following Hamiltonian
\begin{equation}
H_{MF}=\sum_{i,j}\psi_{i}^{\dagger}U_{i,j}\psi_{j}.\nonumber
\end{equation}
Here $\psi_{i}=\left(\begin{array}{c}f_{i,\uparrow}\\f^{\dagger}_{i,\downarrow}\end{array}\right)$, 
$U_{i,j}=\left(\begin{array}{cc}\chi_{i,j} & \Delta^{*}_{i,j} \\\Delta_{i,j} & -\chi^{*}_{i,j}\\ \end{array}\right)$. $\chi_{i,j}$ and $\Delta_{i,j}$ denote the RVB parameter in the hopping and pairing channel. 

We note that the RVB state so constructed is invariant when we perform a $SU(2)$ gauge transformation of the form $U_{i,j}\rightarrow G^{\dagger}_{i}U_{i,j}G_{j}$ on the RVB parameter $U_{i,j}$, in which $G_{i}$ is a site dependent $SU(2)$ matrix\cite{Wen1}. Thus, to generate a symmetric spin liquid state, the RVB order parameter $U_{i,j}$ should be invariant under the symmetry operations only up to a $SU(2)$ gauge transformation. The gauge inequivalent way to choose such a gauge transformation provides a guiding principle to classify the resultant RVB states\cite{Wen1}. For example, in a $Z_{2}$ spin liquid state, the translational symmetry can be realized either by assuming a translational invariant RVB ansatz, or an RVB ansatz that differ by a $Z_{2}$ gauge transformation from the translated ansatz. Here we only consider $Z_{2}$ spin liquid state of the second type, which can have a smooth connection with the $U(1)$ Dirac spin liquid state.

At the same time, we note that although the exchange coupling in the Hamiltonian is restricted to the nearest-neighboring bonds, the RVB parameter $U_{i,j}$ can be more extended in space. In our study, we will keep RVB order parameters $U_{i,j}$ up to the third neighboring bonds. We find that the inclusion of longer range RVB parameter is crucial in the variational study of the spin liquid state in the vicinity of the $U(1)$ Dirac spin liquid state, around which the mapping between the RVB parameters and the spin liquid state is non-injective\cite{Tao2}.

The mean field ansatz of the $U(1)$ and the $Z_{2}$ spin liquid state studied in this work are illustrated in Figure 1. The green parallelogram donotes the unit cell of the Kagome lattice, with $\mathrm{\mathbf{a}}_{1}$ and $\mathrm{\mathbf{a}}_{2}$ as its two basis vectors. The blue, yellow and pink lines denote the first, second and the third neighboring RVB parameters $U_{i,j}$. RVB parameters in other unit cells can be found from them 
through lattice translation. For the spin liquid state studied here, $U_{i,j}$ is translational invariant along the $\mathrm{\mathbf{a}}_{2}$ direction, but will change sign when translated in the $\mathrm{\mathbf{a}}_{1}$ direction by one lattice constant, if the cell index of site $i$ and $j$ in the $\mathrm{\mathbf{a}}_{2}$ direction differ by an odd number.

\begin{figure}
\includegraphics[width=7cm,angle=0]{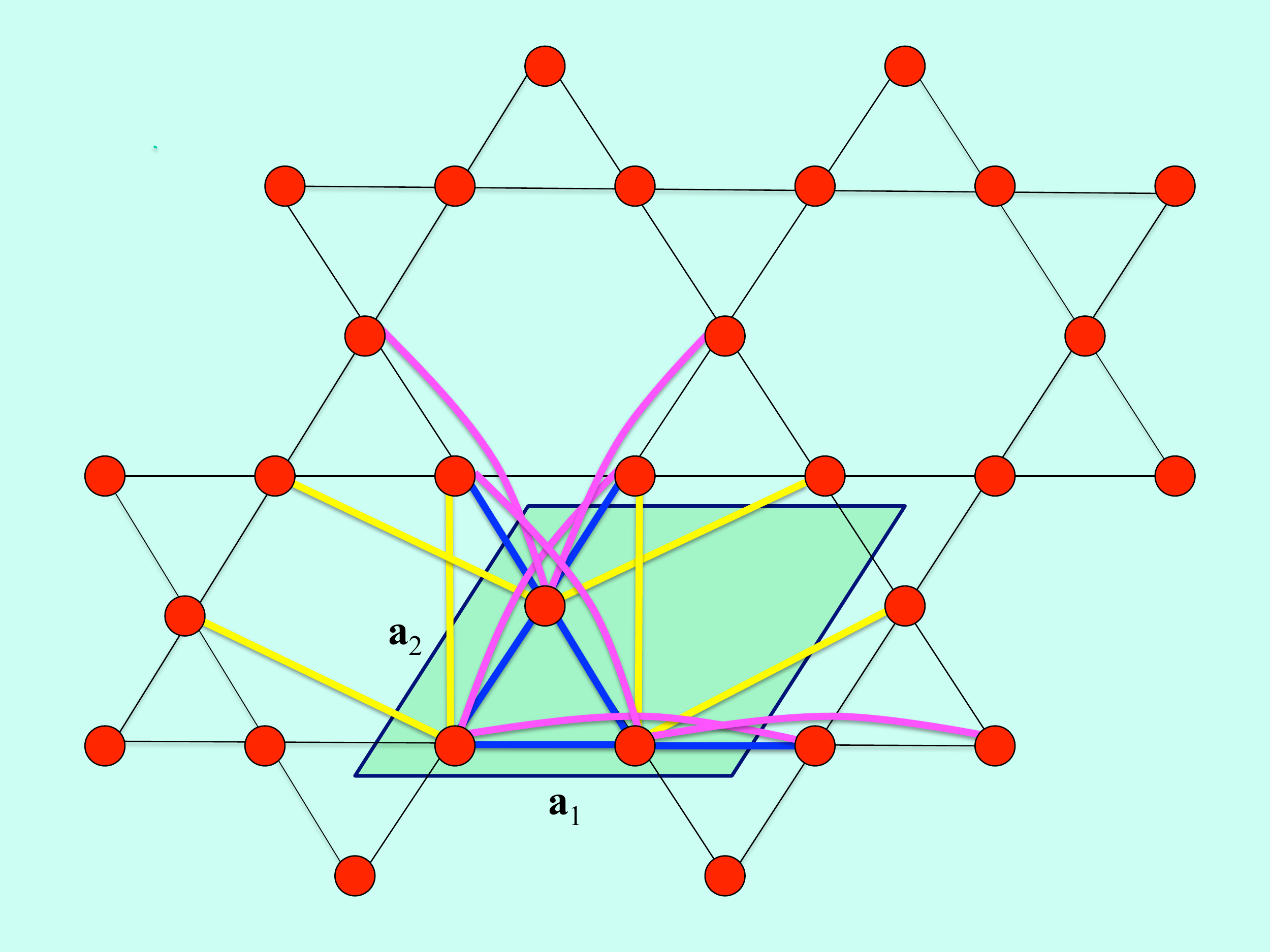}
\caption{Illustration of the mean field ansatz of the $U(1)$ and the $Z_{2}$ spin liquid state studied in this paper. The green 
parallelogram denotes the unit cell of the Kagome lattice, with $\mathrm{\mathbf{a}}_{1}$ and $\mathrm{\mathbf{a}}_{2}$ as two basis vectors.
The blue, yellow and pink lines denote the first, second and the third neighboring RVB parameters $U_{i,j}$. RVB parameters in other unit cells can be generated from them
through lattice translation. Note that $U_{i,j}$ is translational invariant along the $\mathrm{\mathbf{a}}_{2}$ direction, but will change sign when translated in the $\mathrm{\mathbf{a}}_{1}$ direction by one lattice constant, if the cell indices in the $\mathrm{\mathbf{a}}_{2}$ direction of site $i$ and $j$ differ by an odd number. } \label{fig1}
\end{figure}

For the $U(1)$ spin liquid state, the RVB parameters take the form
\begin{eqnarray}
U_{i,j}=\left\{\begin{aligned}
                -s_{i,j}\ \tau_{3}& & \mathrm{first\ neighbor}\\  
                -s_{i,j}\ \rho\tau_{3}& & \mathrm{second\ neighbor}\\
                -s_{i,j}\ \eta\tau_{3}& & \mathrm{third\ neighbor}
\end{aligned}  
\right.
\end{eqnarray}
Here a chemical potential term is implicitly assumed to enforce the half-filling condition on the Fermion number. $\tau_{3}=\left(\begin{array}{cc}1 & 0 \\0 & -1\end{array}\right)$ is a Pauli matrix and $\rho,\eta$ are two real variational parametrs. $s_{i,j}=\pm1$ is introduced to generate the sign change when we translate $U_{i,j}$ in the $\mathrm{\mathbf{a}}_{1}$ direction. They equal to 1 for the RVB parameters illustrated in Figure 1. In a previous work\cite{Tao2}, we have shown that for $-0.6\leq\rho=\eta\leq0.2$, the ground state of the mean field Hamiltonian is independent of $\rho$. 

For the $Z_{2}$ spin liquid state, the RVB parameters take the form
\begin{eqnarray}
U_{i,j}=\left\{\begin{aligned}
                -\mu \vec{n}_{\phi_{1}} \cdot \vec{\tau} & & \mathrm{on-site} \\
                -s_{i,j}\ \tau_{3}& & \mathrm{first\ neighbor}\\  
                -s_{i,j}\ \rho  \vec{n}_{\phi_{2}} \cdot \vec{\tau}& & \mathrm{second\ neighbor}\\
                -s_{i,j}\ \eta  \vec{n}_{\phi_{3}} \cdot \vec{\tau}& & \mathrm{third\ neighbor}
\end{aligned}  
\right.
\end{eqnarray}
Here $\vec{\tau}=(\tau_{1},\tau_{2},\tau_{3})$ are the Pauli matrices, $\vec{n}_{\phi}=(\sin\phi,0,\cos\phi)$ is a unit vector in the $\tau_{1}-\tau_{3}$ plane.
 $\mu,\rho,\eta$ and $\phi_{1,2,3}$ are six real variational parameters of the $Z_{2}$ spin liquid state. It can be easily checked that spin liquid state generated from the $U(1)$ and the $Z_{2}$ ansatz respect all physical symmetry of the KAFH. At the same time, the $Z_{2}$ spin liquid state reduces to the $U(1)$ spin liquid state when $\phi_{1,2,3}=N\pi$, in which $N$ is an arbitrary integer.

\begin{figure}
\includegraphics[width=8cm,angle=0]{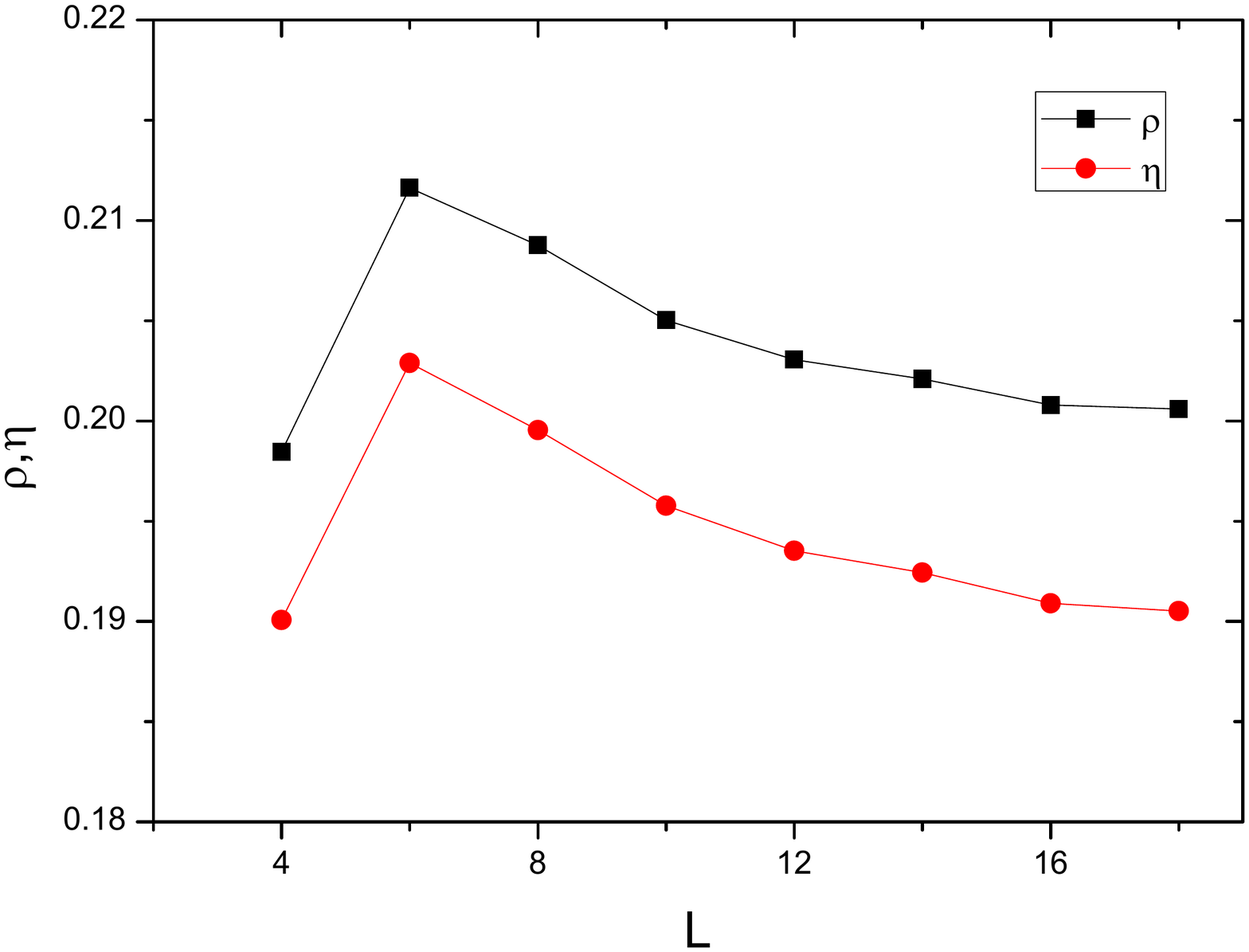}
\caption{The optimized RVB parameters of the $U(1)$ spin liquid state. } \label{fig2}
\end{figure}

\begin{figure}
\includegraphics[width=8cm,angle=0]{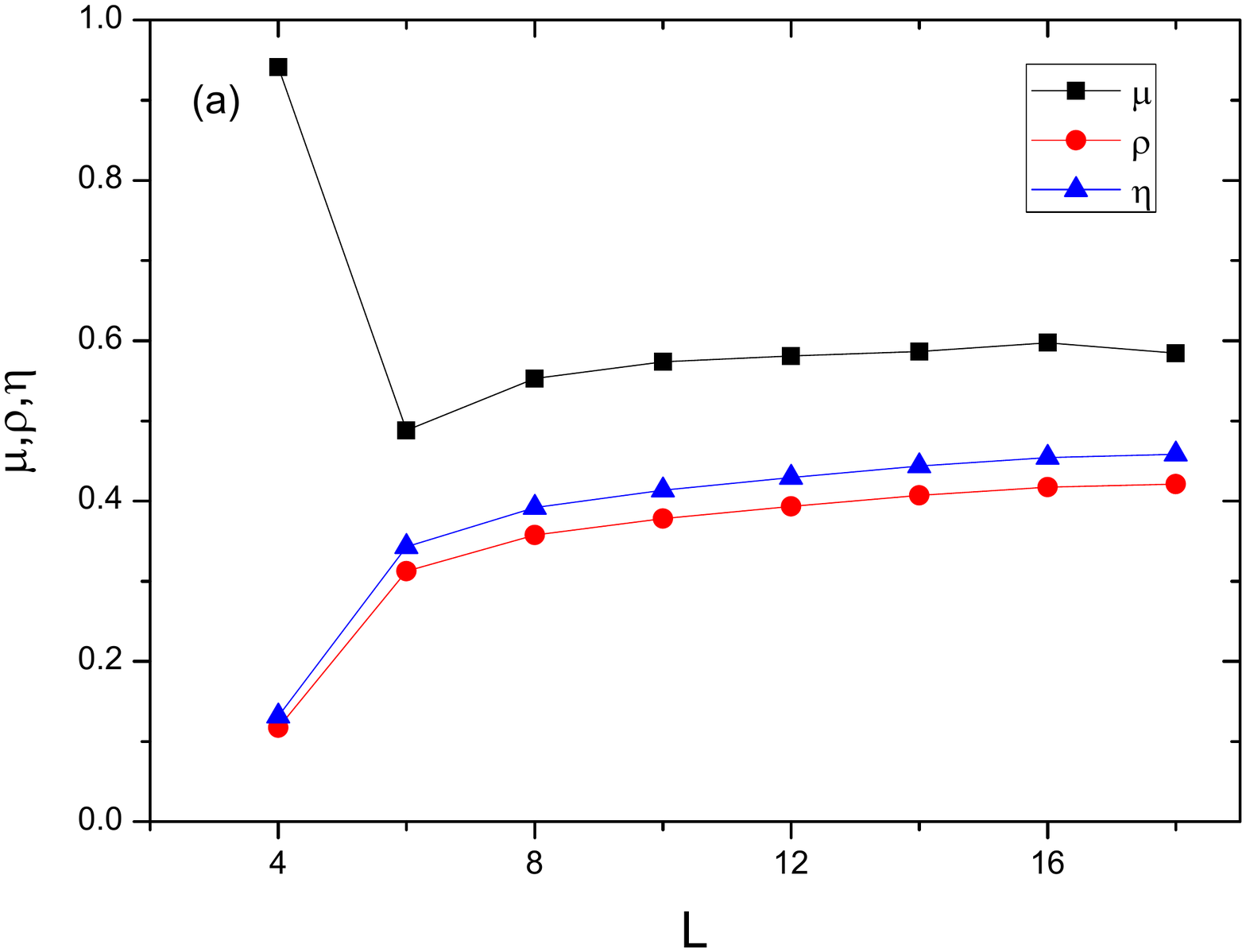}
\includegraphics[width=8cm,angle=0]{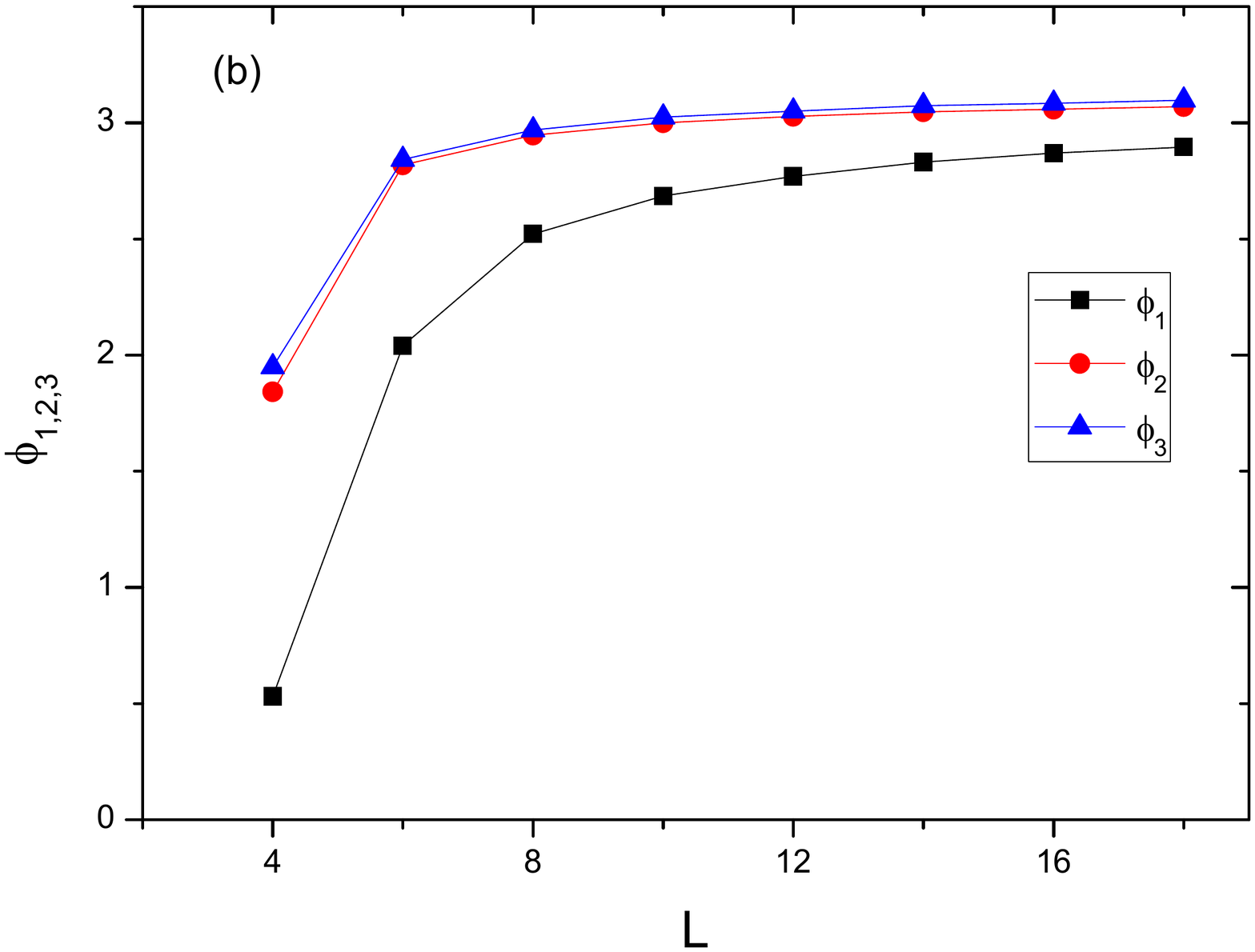}
\caption{The optimized RVB parameters of the $Z_{2}$ spin liquid state, (a)the amplitude $\mu$, $\rho$ and $\eta$, (b)the gauge angle $\phi_{1,2,3}$. } \label{fig3}
\end{figure}

The remaining task is to find the optimized value of the RVB parameters for the spin-$\frac{1}{2}$ NN-KAFH. 
The calculation is done on a $L\times L\times 3$ cluster, with periodic - anti-periodic boundary condition on the mean field ansatz. In a series of previous variational studies\cite{Ran,Iqbal1,Iqbal2,Iqbal3,Iqbal4,Iqbal5}, it is claimed that the $U(1)$ Dirac spin liquid state is the best variational state of the spin-$\frac{1}{2}$ NN-KAFH. Here we show that this is not true. 
In Figure 2 and 3, we plot the optimized values of the RVB parameters for the $U(1)$ and the $Z_{2}$ spin liquid state. The largest cluster size that we have achieved good convergence in the RVB parameters is $L=18$, beyond which the optimization procedure becomes numerically too expensive. The minimum of the $U(1)$ spin liquid state is found to be very close to the line $\rho=\eta$, where the mapping between the $U(1)$ spin liquid state and the RVB parameters is non-injective.

In Figure 4, we plot the energy of the $U(1)$ and the $Z_{2}$ spin liquid state for cluster with $L=4$ through that with $L=18$. The results show convincingly that the $Z_{2}$ spin liquid state is more stable than the $U(1)$ spin liquid state in the thermodynamic limit. We find that the energy of the $U(1)$ and the $Z_{2}$ spin liquid state exhibit different scaling behaviors at large $L$. In particular, while the energy of the $Z_{2}$ spin liquid state increases monotonically with $L$ as $E_{L}=E_{\infty}-\alpha L^{-4}$ for large $L$, a maximum appears at $L=12$ in the energy of the $U(1)$ spin liquid state. Thus, a simple polynomial extrapolation of the energy difference between the $U(1)$ and the $Z_{2}$ spin liquid state from the data with small $L$, as was done in Ref.[\onlinecite{Iqbal4}], can not provide useful information concerning the thermodynamic limit.

\begin{figure}
\includegraphics[width=8cm,angle=0]{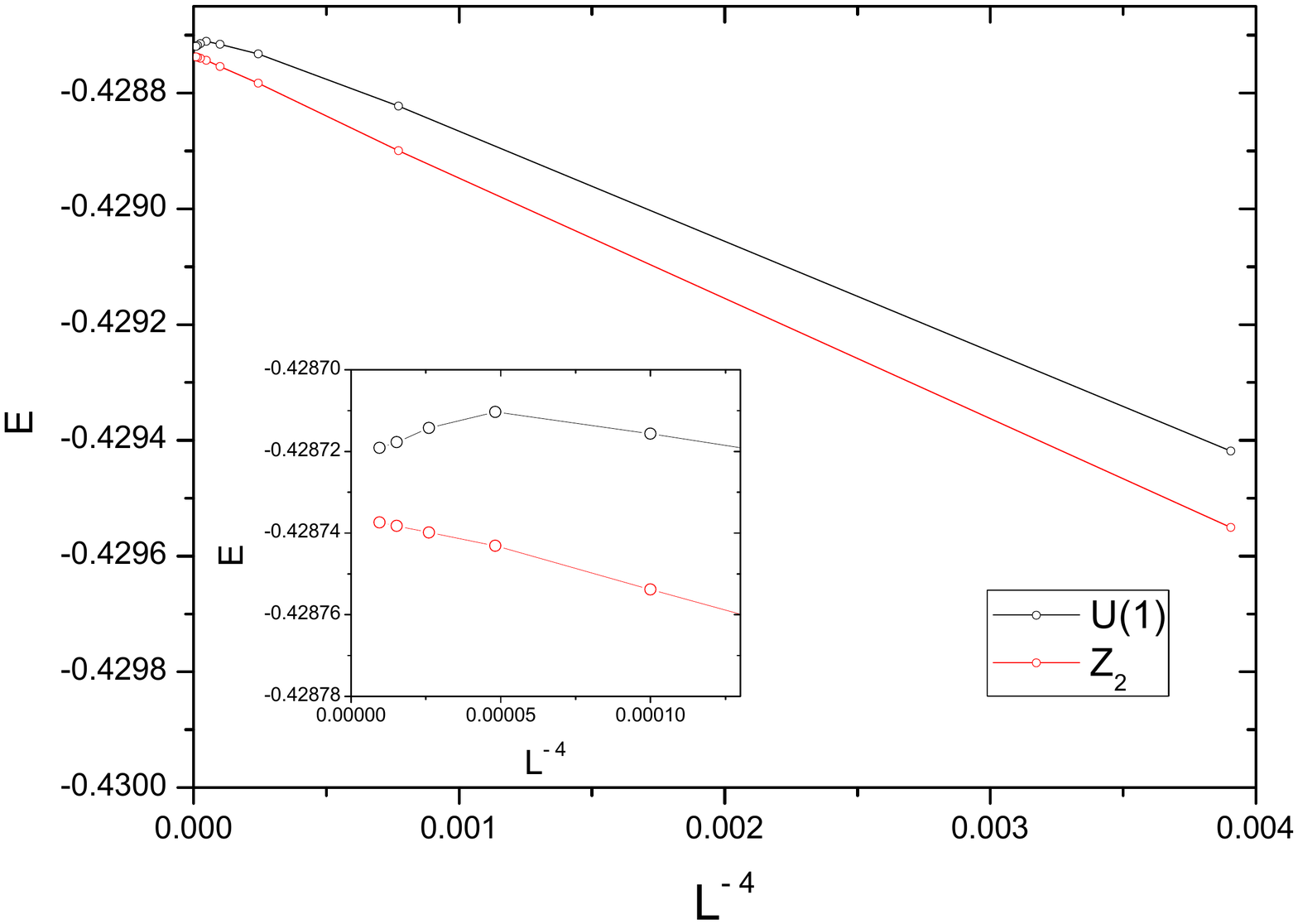}
\caption{The variational energy of the $U(1)$ and the $Z_{2}$ spin liquid state. The inset shows the details for large $L$. The error bar of the data is smaller than the symbol size.}
\end{figure}

We now show that the $Z_{2}$ spin liquid state we find is indeed gapped. In principle, it is impossible to extract the excitation gap directly from the variational ground state. Here we will present instead the result of the mean field spinon gap. The amplitude of the RVB parameter on the nearest-neighboring bonds will be taken as the unit of energy. The spinon gap so defined provides a relative measure of the spin excitation gap in terms of the spinon band width. The spinon gap on finite clusters for both the $U(1)$ and the $Z_{2}$ spin liquid state are plotted in Figure 5. We find that the spinon gap in the $U(1)$ spin liquid state extrapolates to zero in the thermodynamic limit following the scaling $\Delta\approx \alpha L^{-1}$, which is the expected behavior for a Dirac spin liquid. On the other hand, the spinon gap of the $Z_{2}$ spin liquid state is found to approach a finite value of $\Delta_{0}\simeq0.1$ in the thermodynamic limit with the scaling $\Delta\approx \Delta_{0}+\beta L^{-2}$, as is expected for a gapped system\cite{Sorella}. 

\begin{figure}
\includegraphics[width=8cm,angle=0]{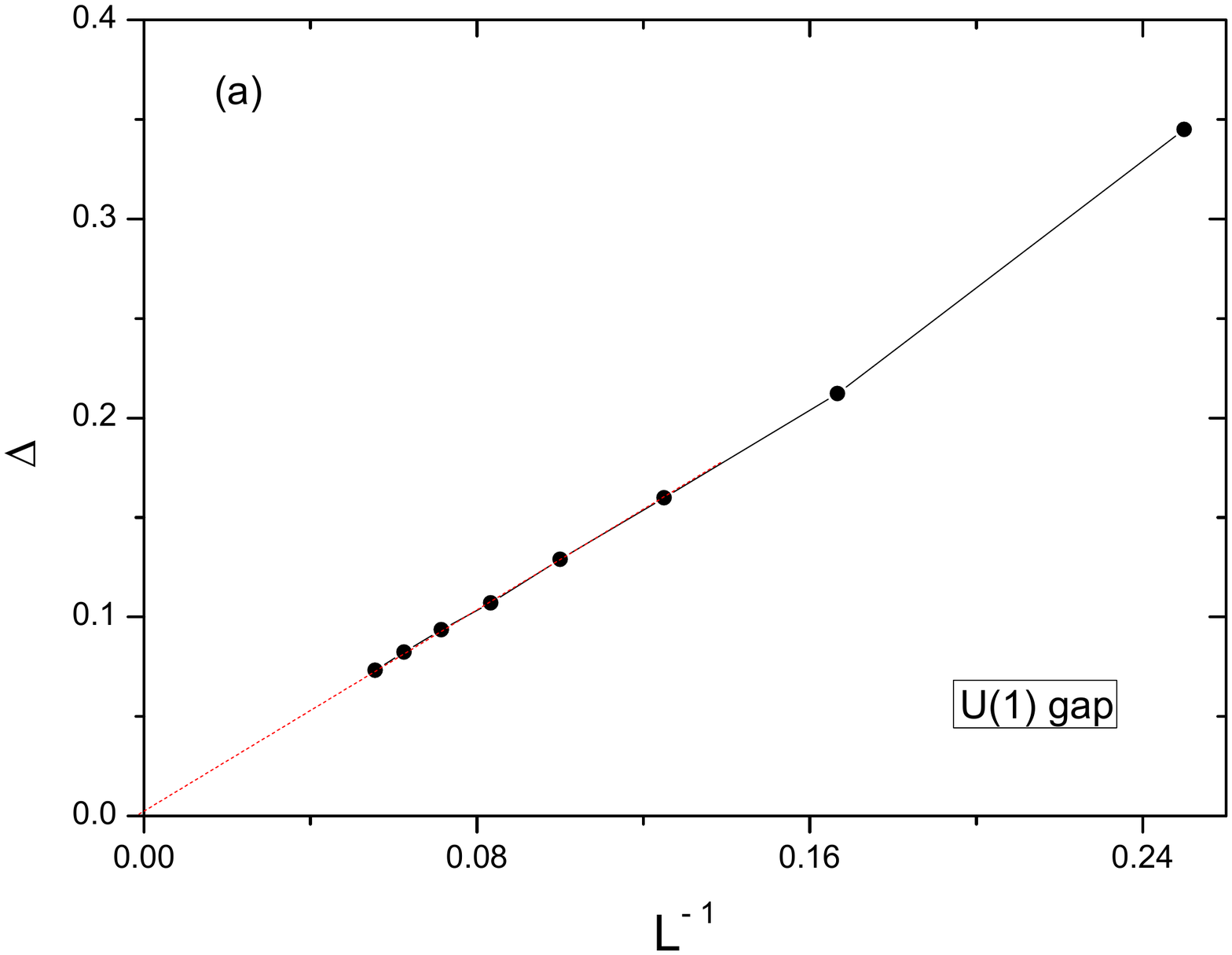}
\includegraphics[width=8cm,angle=0]{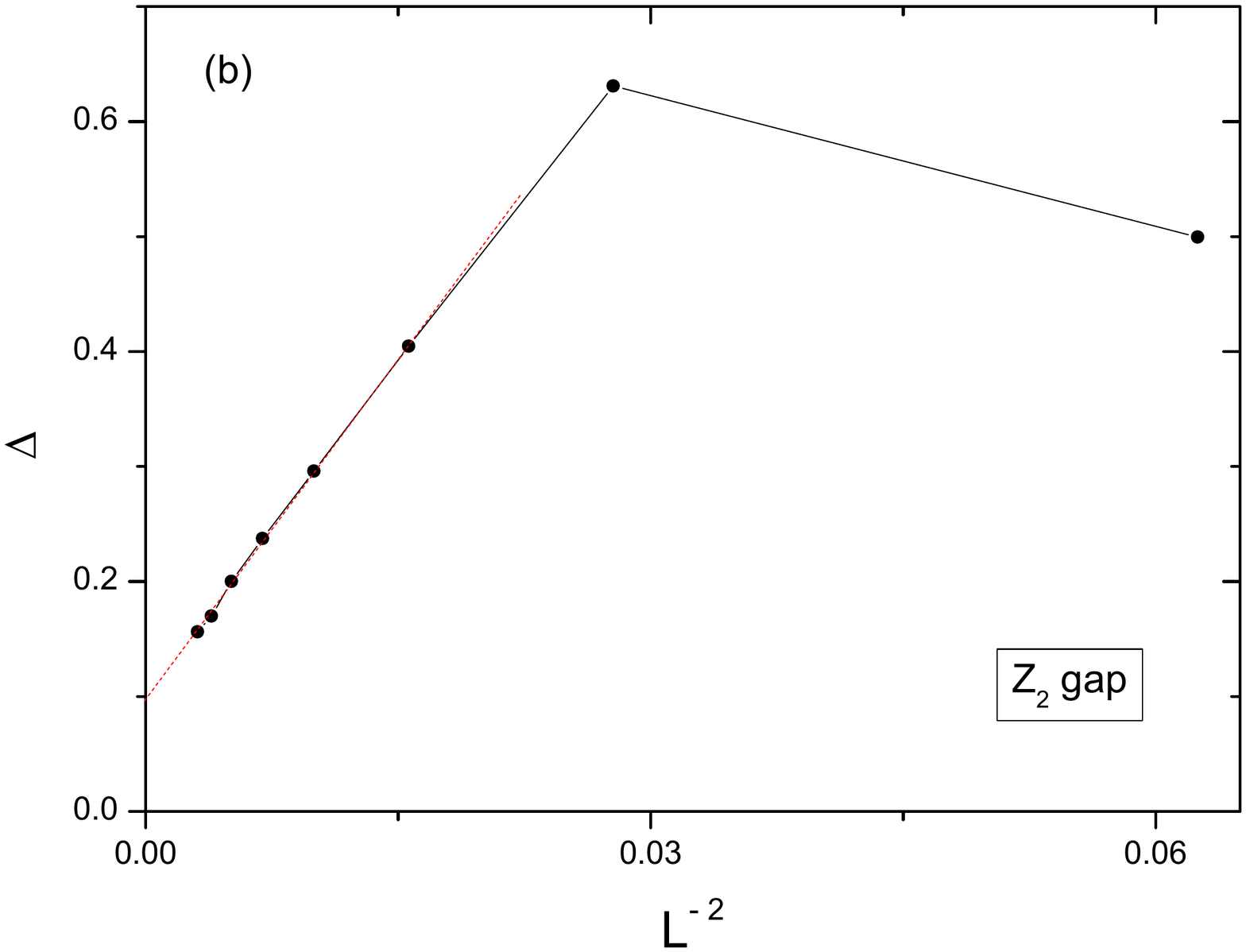}
\caption{The scaling of the mean field spinon gap with $L$. The red dashed lines denote the linear fitting of the data. Note that the spinon gap in the $U(1)$ and the $Z_{2}$ spin liquid state exhibit different scaling with $L$.}
\end{figure}

We note that the energy difference between the $U(1)$ and the $Z_{2}$ spin liquid state is extremely small, although they have very different RVB parameters.
To find if the two states are close to each other in the Hilbert space, we have calculated the overlap between them. We find that the overlap is quite large on finite clusters. For example, the overlap is as high as $0.93$ on a $L=18$ cluster, which is the largest one on which we have achieved good convergence in the RVB parameters. The exceptional insensitivity of the RVB state to the change in the RVB parameters as exposed here is rather unusual and may be attributed to the same flat band physics discussed in Ref.[\onlinecite{Tao2}]. Indeed, one find that the second and the third neighboring RVB parameters are very close to each other in both the $U(1)$ and the $Z_{2}$ spin liquid state.

 \begin{figure}
\includegraphics[width=8cm,angle=0]{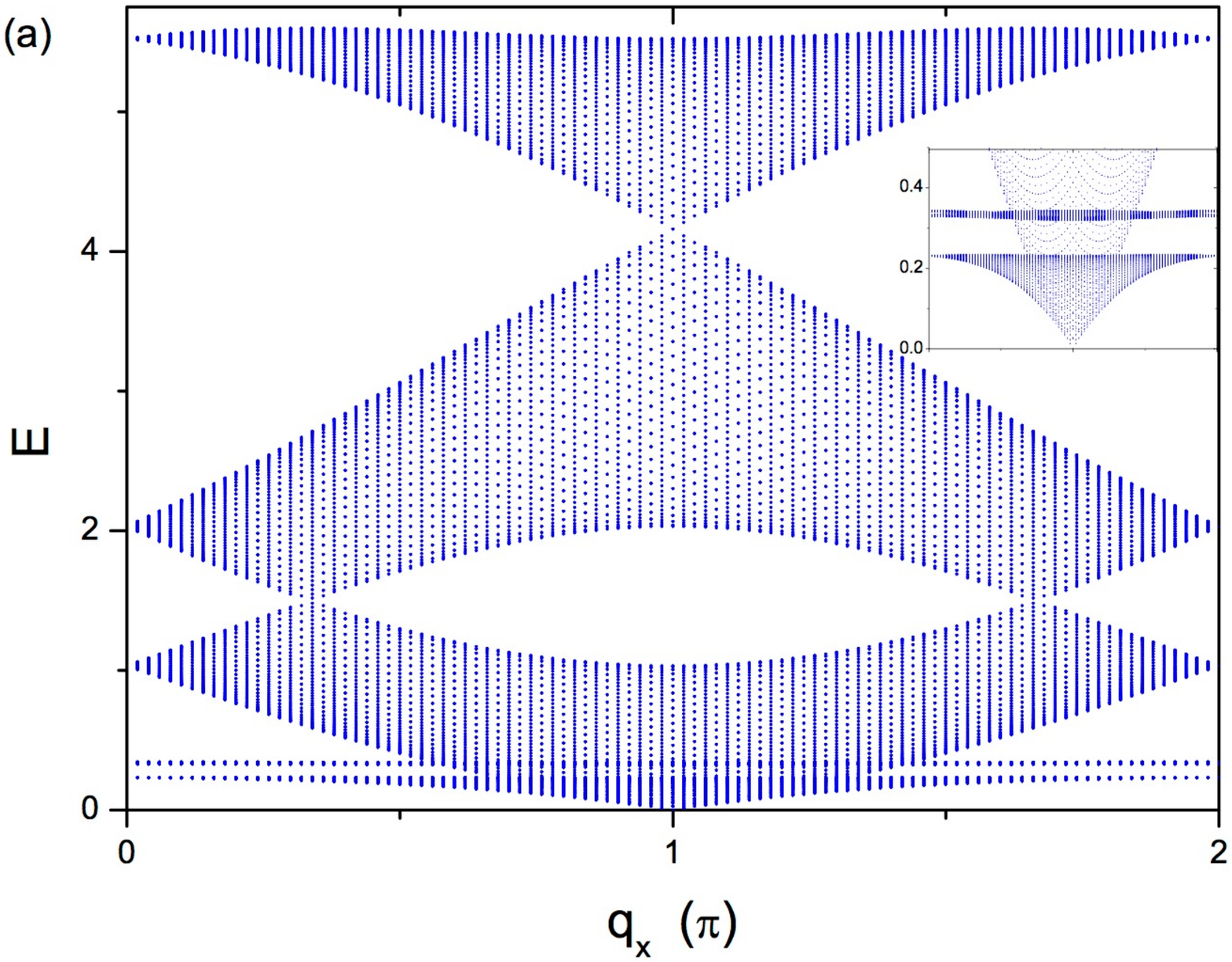}
\includegraphics[width=8cm,angle=0]{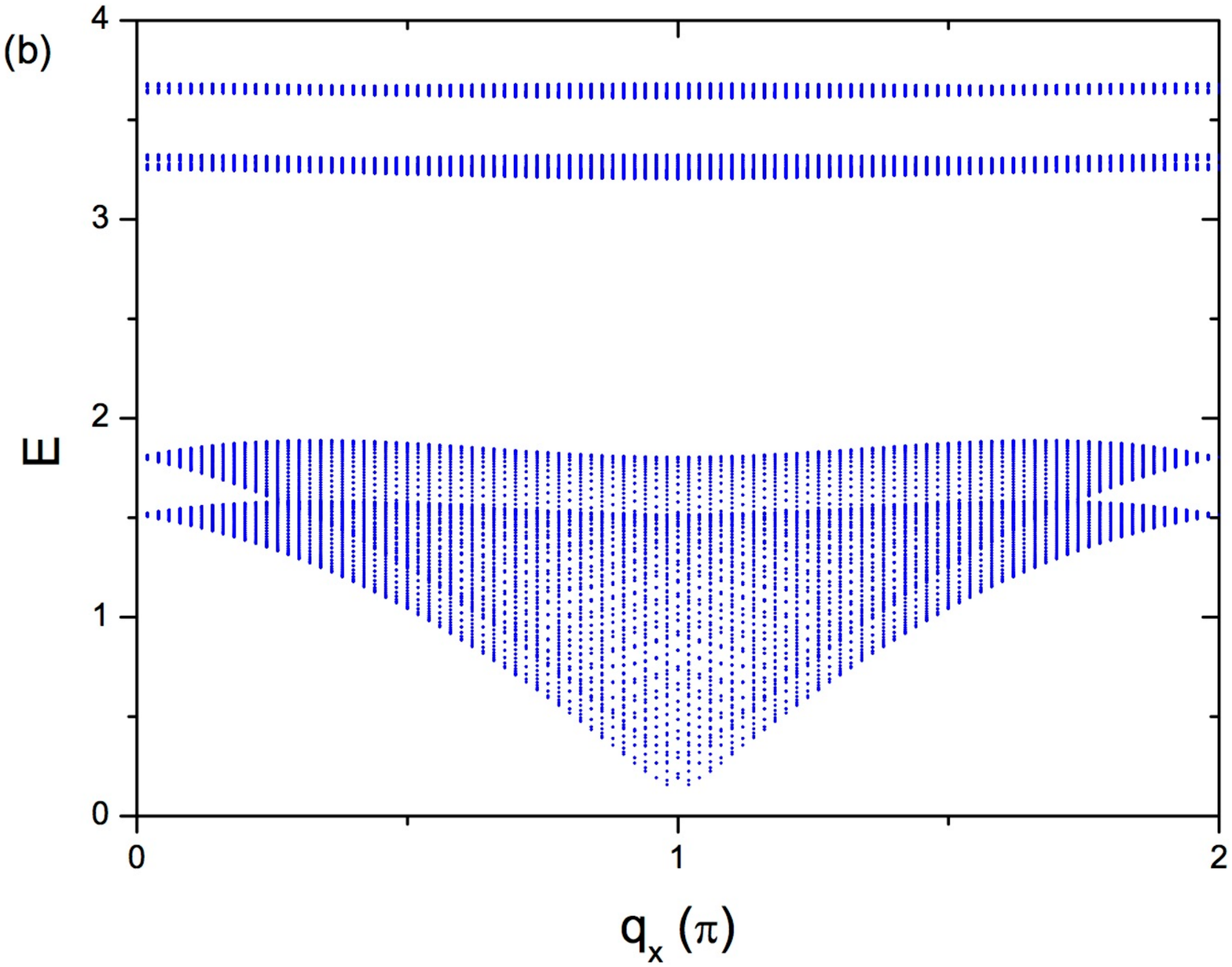}
\caption{(a) and (b) The mean field spinon spectrum of the $U(1)$ and the $Z_{2}$ spin liquid state. The inset of (a) presents the low energy details of the spinon spectrum of the $U(1)$ spin liquid state. Both spectrums are calculated with the optimized RVB parameters for $L=18$. In the case of the $U(1)$ spin liquid, we have folded the hole-side of the spinon spectrum to positive energy so as to facilitate a meaningful comparison with the spectrum of the $Z_{2}$ spin liquid state.}
\end{figure}  
 
On the other hand, the spinon spectrum of the $U(1)$ and the $Z_{2}$ spin liquid state, which are plotted in Figure 6(a) and 6(b), are totally different. In particular, while the quasi-flat band appears at rather low energy in the spectrum of the $U(1)$ spin liquid state, they are pushed to the top of the spectrum in the $Z_{2}$ spin liquid state. The extreme insensitivity of the ground state energy to the excitation spectrum as exposed here clearly demonstrates the virtue of the RVB theory, which provides not only a variational understanding of the ground state structure, but also a comprehensive picture for the excitation spectrum of the system. Such spectroscopic information would be inaccessible with other numerical approaches that focus only on the ground state. We note that except for the small gap, the spinon excitation spectrum of the $Z_{2}$ spin liquid state is very close to that of a standard Dirac spin liquid state with a particle-hole symmetric spinon excitation spectrum. It is thus not surprising to find Dirac-like spinon response on a finite cluster\cite{He}.

In conclusion, we find that a $Z_{2}$ spin liquid state is more stable than the generally believed $U(1)$ Dirac spin liquid state for the spin-$\frac{1}{2}$ NN-KAFH. We find that the $Z_{2}$ spin liquid state has a Dirac-like spinon excitation spectrum with a small gap of about $1/40$ of the spinon band width. We also find that while the $Z_{2}$ spin and the $U(1)$ spin liquid state have a rather large overlap on finite clusters, they host totally different spinon excitation spectrums. The small size of the spinon gap and the large overlap of the ground state with the $U(1)$ Dirac spin liquid state on finite clusters indicate that the spin-$\frac{1}{2}$ NN-KAFH should be better understood as a nearly critical system, rather than a prototypical gapped $Z_{2}$ spin liquid system\cite{Tao3}. We believe this is the reason why the ground state of the spin-$\frac{1}{2}$ NN-KAFH is so controversial. It is interesting to see if one can drive the system deeper into the gapped $Z_{2}$ spin liquid phase by introducing perturbation away from the NN-KAFH point, so that one can conclude the nature of the spin liquid state with more confidence in numerical simulations. The fully frustrated perturbation proposed in Ref.[\onlinecite{Tao2}] is particularly promising in this respect.

We acknowledge the support from the grant NSFC 11674391 and the Research Funds of Renmin University of China.

\end{document}